\documentclass[useAMS,usenatbib,letterpaper]{mn2e}
\usepackage[totalwidth=480pt, totalheight=680pt]{geometry}
\usepackage{graphicx}

\newcommand{\ra}{\mbox{$\alpha_{\hbox{J2000}}$}}
\newcommand{\dec}{\mbox{$\delta_{\hbox{J2000}}$}}
\newcommand{\dg}{\mbox{$^\circ$}}
\newcommand{\hr}{\mbox{$^{\rm h}$}}
\newcommand{\am}{\mbox{$^{\prime}$}}

\newcommand{\as}{\mbox{$^{\prime\prime}$}}

\newcommand{\dsc}{$\delta$ Scuti~}
\newcommand{\mdot}{\mbox{$M_{\odot}$}}
\newcommand{\rdot}{\mbox{$R_{\odot}$}}
\newcommand{\ldot}{\mbox{$L_{\odot}$}}
\newcommand{\mbol}{\mbox{$M_{bol}$}}

\newcommand{\mnras}{MNRAS}

\newcommand{\aaa}{A\&A}

\newcommand{\aas}{A\&AS}

\title[The first high-amplitude delta Scuti star in an eclipsing binary system]{The first high-amplitude delta Scuti star in an eclipsing binary system}
\author[J. L. Christiansen et al.]{
	J. L. Christiansen$^{1}$\thanks{E-mail:jessiec@phys.unsw.edu.au (JLC)}, 
	A. Derekas$^{2}$, 
	M. C. B. Ashley$^{1}$, 
	J. K. Webb$^{1}$,
 \newauthor	 
	M. G. Hidas$^{1,3,4}$, 
	D. W. Hamacher$^{1}$ 
	and L. L. Kiss$^{2}$\\
$^{1}$School of Physics, University of New South Wales, Sydney, 2052, Australia\\
$^{2}$School of Physics, University of Sydney, Sydney, 2006, Australia\\
$^{3}$Las Cumbres Observatory Global Telescope, Goleta, CA 93117, USA\\
$^{4}$Department of Physics, University of California, Santa Barbara, CA 93106, USA}
\begin{document}

\date{\today}

\pagerange{\pageref{firstpage}--\pageref{lastpage}} \pubyear{2007}

\maketitle

\label{firstpage}

\begin{abstract}
We report the discovery of the first high-amplitude \dsc star in an eclipsing binary, which we have designated UNSW-V-500. The system is an Algol-type semi-detached eclipsing binary of maximum brightness $V = 12.52$ mag. A best-fitting solution to the binary light curve and two radial velocity curves is derived using the Wilson-Devinney code. We identify a late A spectral type primary component of mass $1.49\pm0.02$ \mdot~and a late K spectral type secondary of mass $0.33\pm0.02$ \mdot, with an inclination of $86.5\pm1.0^{\circ}$, and a period of $5.3504751\pm0.0000006$ d. A Fourier analysis of the residuals from this solution is performed using {\small PERIOD04} to investigate the \dsc pulsations. We detect a single pulsation frequency of $f_1 = 13.621\pm0.015$ cd$^{-1}$, and it appears this is the first overtone radial mode frequency. This system provides the first opportunity to measure the dynamical mass for a star of this variable type; previously, masses have been derived from stellar evolution and pulsation models.
\end{abstract}


\section{Introduction}
\label{sec:int}

The serendipitous arrangement of a pulsating star in an eclipsing binary system represents a unique laboratory for astrophysical measurements. The binarity constrains the physical and geometrical parameters of the system, and can also assist in mode identification in the pulsations. Since the pulsating star and non-pulsating companion can reasonably be assumed to have formed from the same parent cloud, we can utilise information from the non-pulsating companion in identifying stellar evolution models for pulsating stars. We present here the first known example of a high-amplitude \dsc (HADS) star in an eclipsing binary system, designated UNSW-V-500.

\begin{figure}
\begin{center}
\includegraphics[width=8cm]{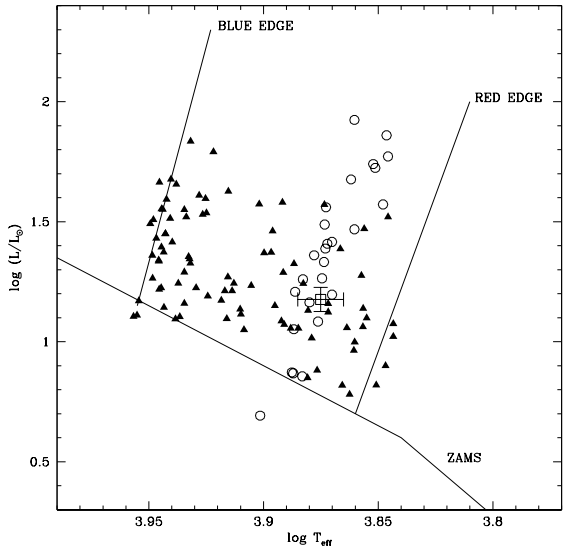}
\caption{\small  A HR diagram showing the positions of $\delta$ Scuti stars in eclipsing binaries. The solid triangles are data from Table 4 of \protect \cite{Soydugan06}. The open circles are high-amplitude \dsc stars from Table 2 of \protect \cite{McNamara00b}. UNSW-V-500 is shown as an open square. The observational red and blue edges of the classical instability strip are shown, as well as the theoretical zero-age main sequence.}
\label{fig:HR}
\end{center}
\end{figure}

\dsc stars are the main-sequence analogues of Cepheid variables. They are late A to early F spectral type, and pulsate with periods of between 1--6 hours. They are typically on or slightly above the main-sequence. Low-amplitude \dsc stars typically pulsate in many higher, non-radial modes simultaneously with amplitudes of less than 0.05 mag. High-amplitude \dsc pulsate primarily in the radial modes and have higher amplitudes, with the conventional cut-off given as $A_{\rm V}\geq0.30$ mag. Fig. \ref{fig:HR} shows the \dsc region of the HR diagram; high-amplitude \dsc stars are constrained to a narrower range in $T_{\rm eff}$ of width $\sim$300 K within this region. The subset of low metallicity Population II HADS stars are designated as SX Phe stars. Prior to the discovery of UNSW-V-500, all bright field HADS stars were identified as pulsating in the fundamental radial mode. A significant fraction ($\sim$40 per cent) are double-mode pulsators, additionally pulsating in the radial first overtone mode \citep{McNamara00a}. However, SX Phe stars have been identified in globular clusters as pulsating in the first overtone rather than the fundamental mode \citep{McNamara00b,Nemec94}.

\cite{Soydugan06} present a list of 25 confirmed eclipsing binary systems with pulsating components in the \dsc region of the instability strip. All of these have low pulsation amplitudes, ranging from $A_{\rm V} =$ 0.007--0.02 mag and up to $A_{\rm B} =$ 0.06 mag. Three systems that have been studied in detail are Y Cam \citep{Kim02a}, AS Eri \citep{Mkrtichian04} and AB Cas \citep{Rodriguez04}. Parameters of these systems are shown in Table \ref{tab:ebs}, with the parameters of UNSW-V-500 shown for comparison.

We note that although 35 \dsc stars in eclipsing binary systems have been identified (25 in \cite{Soydugan06}; 9 recently in \cite{Pigulski07}; and the subject of this paper), UNSW-V-500 is the first HADS star. This is a curious statistic, since with larger amplitudes, surveys might be expected to be observationally biased towards finding HADS stars in these systems. A census of the Rodr{\' i}guez, L{\' o}pex-Gonz{\' a}lez \& L{\'o}pez de Coca (2000) catalogue shows that the detected fraction of HADS stars is $\sim$25 per cent of the total \dsc population. One assumes that this fraction is influenced by the same observational selection bias affecting the detection of HADS stars in eclipsing binary systems; we are seeing a detection rate nearly an order of magnitude below this. However, it is difficult to draw conclusions from the small numbers of binary systems that have been found.

\begin{table*}
 \centering
 \begin{minipage}{140mm}
 \label{tab:ebs}
  \caption{A sample of \dsc stars in eclipsing binary systems. A$_{\rm V}$ is the amplitude of the \dsc pulsations.}
  \begin{tabular}{@{}lcccccc@{}}
  \hline
   Name                    &  V                  & A$_{\rm V}$ & Spectral Type & P$_{\rm puls}$ & P$_{\rm orb}$ & Inclination \\
                                  &  mag             & mag                     &                           & (d)          & (d)        & $^\circ$       \\
 \hline 
 Y Cam                      & 10.56           & 0.04                     & A7V                   & 0.063           & 3.3055      &  86               \\
 AB Cas                     &  10.16          & 0.05                     & A3V                   & 0.058           & 1.3669      &  87.5           \\
 AS Eri                       &   8.31  	     &  0.0068               & A3V                   &  0.016          & 2.6642      &  --          \\
 UNSW-V-500 &  12.52\footnote{V$_{\rm max}$ from the ASAS catalogue.}  &  $\sim$0.35$\pm$0.5        & A7V               &  0.0734$\pm$0.0001          & 5.3504751$\pm$0.0000006  & 86.5$\pm$1.0\\
\hline
\end{tabular}
\end{minipage}
\end{table*}

\section[]{Observations}
\label{sec:obs}

\subsection[]{Photometry}
\label{sec:phot}

UNSW-V-500 was initially observed over 29 nights from February 2006 to April 2006. The photometric {\it I\/}-band observations were performed with the 0.5-m Automated Patrol Telescope at Siding Spring Observatory, Australia. The observations formed part of an extrasolar planet transit search being undertaken by the University of New South Wales \citep{Hidas05}. The CCD has 770 $\times$ 1150 pixels and images a 2 $\times$ 3 deg$^2$ field with a relatively low spatial resolution of 9.4 arcsec per pixel. We have used a customised aperture photometry data reduction pipeline to construct our target light curves. The results of the project include the discovery of a new eclipsing system of K7 dwarf components \citep{Young06}.

Identification of UNSW-V-500 as demonstrating both eclipsing and pulsating variations was made during routine cataloguing of the variable light curves detected in the transit search, to be published separately. The photometry aperture used in the reduction process, which is nearly 1 arcmin in diameter, can usually be expected to contain more than one star due to crowding effects in the target fields (typically chosen to have a Galactic latitude of 10--20\dg). Higher resolution images from the Digitized Sky Survey catalogue showed that in the case of UNSW-V-500 the photometry aperture contained one bright central star and 6 additional stars at least 3.5 magnitudes fainter. Due to the large amplitude of the \dsc pulsations (diluted to $\sim$0.1 mag in the original photometry aperture) the system was identified with the bright central star, \ra~=~13\hr~10\am~18\farcs7, \dec~=~$-45$\dg~9\am~13\as. A catalogue search revealed that this system had been previously observed and identified in the All Sky Automated Survey Catalog of Variable Stars III \citep{Pojmanski04} as an eclipsing binary system, designated ASAS131018-4509.2. From their data they measured an initial epoch of $T_0 = {\rm JD}2451892.6$ and a period of $P = 5.350479$ d. However, their precision was insufficient for detection of the \dsc pulsation.

\begin{figure}
\begin{center}
\includegraphics[width=8cm]{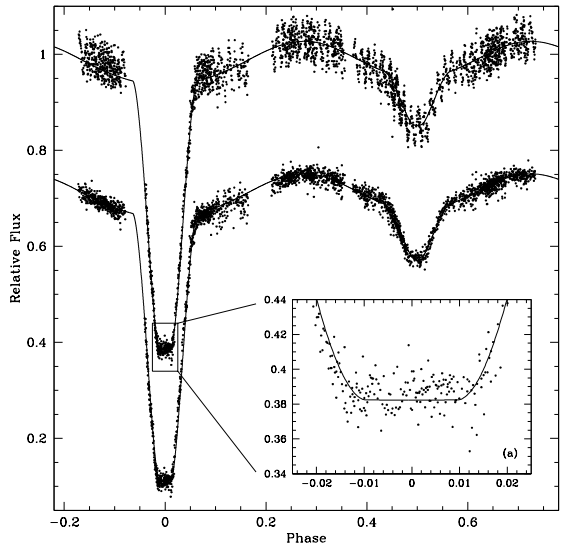}
\caption{\small The phased light curve of 24 nights of data taken with the Automated Patrol Telescope in the $I$ band. The upper curve is the original data, with the scatter outside the primary eclipse due to the \dsc pulsation. The lower curve is the same data with the \dsc pulsation reconstructed from the frequency analysis and removed. In both cases, the solid line is the fit to the original curve using the Wilson-Devinney code. Panel (a) shows the primary eclipse in more detail --- there is no evidence of \dsc variations in this region.}
\label{fig:lc}
\end{center}
\end{figure}

In the original run of 29 nights one secondary minimum and two partial secondary eclipses were observed, but only two primary eclipse egresses. In order to improve coverage of this part of the light curve, UNSW-V-500 was observed again in the same configuration as described previously on two nights in February 2007 and March 2007 at the predicted times of primary eclipse. These two nights and the 22 best nights of the 2006 data are shown in Fig.~\ref{fig:lc}, phased at a period of 5.3504751 d. The primary eclipse data conclusively confirmed that the light curve was not the result of a \dsc star blended with a background eclipsing binary: the flat-bottomed eclipses show no sign of the pulsations that are evident in the remainder of the light curve. Therefore, the \dsc star must be fully eclipsed by the secondary component of the binary during primary eclipse. The inset in Fig.~\ref{fig:lc} shows the flat primary eclipse in more detail.

\begin{figure}
\begin{center}
\includegraphics[angle=-90,width=8cm]{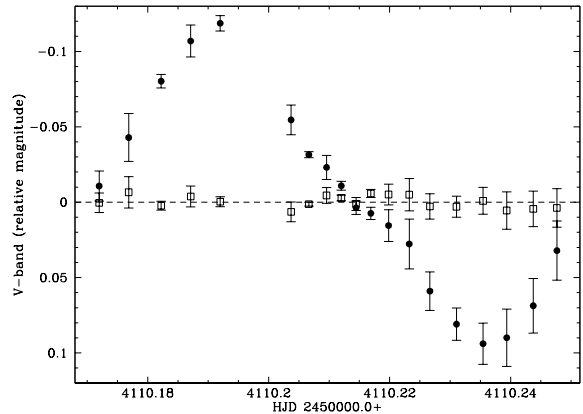}
\caption{\small Light Curves observed with the higher spatial resolution of the 40\as~telescope. The solid circles show the \dsc pulsation of our target, and the open squares show the light curve of the nearby star GSC0824700373 for comparison.}
\label{fig:40lc}
\end{center}
\end{figure}

In order to confirm the identification of the bright central star in the original photometry aperture as the eclipsing binary, higher spatial resolution observations were obtained with the 40-inch telescope at Siding Spring Observatory. The WFI CCD mosaic was used, with an image scale of 0.38 arcsec per pixel. Observations were taken on a single night in January 2007, in the Johnson {\it V\/} filter. These data were reduced using a modified version of our aperture photometry pipeline. The identification of the pulsating star was confirmed and the data are shown in Fig.~\ref{fig:40lc} as solid circles. For comparison, the light curve of a nearby star of similar magnitude, GSC0824700373, is shown as open squares. The lower limit on the amplitude of the \dsc pulsation, diluted in these data by light from the secondary, is $A_{\rm V} = 0.21\pm0.02$ mag. Combined with a primary eclipse depth in the $I$ band of $\sim$60 per cent, the final lower limit is $A_{\rm V} = 0.35\pm0.05$ mag, confirming this star as a high-amplitude \dsc star.

\subsection[]{Spectroscopy}
\label{sec:spec}

Several medium resolution spectra (R $\sim$ 6000) were obtained over two nights in February 2007 with the Double-Beam Spectrograph on the 2.3-m telescope at Siding Spring Observatory. The wavelength range covered was 3900--4400~\AA~in the blue arm and 8000--8900~\AA~in the red. The spectra were reduced using standard {\small IRAF}\footnote{IRAF is distributed by the National Optical Astronomy Observatories, which are operated by the Association of Universities for Research in Astronomy, Inc., under cooperative agreement with the National Science Foundation.} spectroscopy routines. The observations were alternated with arc spectra of FeAr in the blue and NeAr in the red. The flux calibration of the system was performed using the standard stars HR4469 and HR4963. The spectra were rebinned to a resolution of 10~\AA~using the {\small IRAF} routine {\small {\tt REBIN}} and compared with UVILIB template spectra \citep{Pickles98}. Visual inspection resulted in the classification of the spectra as an A7V star. The phase coverage was insufficient to measure the dynamical mass.

To obtain sufficient phase coverage, we obtained additional spectra with the same instrument on five nights in May 2007. The gratings and grating angle were changed to give a wavelength coverage of 3600--4700~\AA~in the blue arm and 6000--7000~\AA~in the red. The same procedure of alternating observations with arc spectra for wavelength calibration was followed, however to improve the stability of the calibration between arc spectra we used the night sky lines present in each spectrum in the red half of the data for additional calibration. The data were then continuum normalised. The red data clearly show spectral features of both the primary and secondary components, and also a significant component of $H\alpha$ emission. Given the semi-detached nature of UNSW-V-500 (see Sec.~\ref{sec:eb}) this may indicate the existence of a gas stream between the two components (see for instance the set of well observed Algol-type eclipsing binaries in \cite{Richards99,Vesper01}). The blue data are entirely dominated by the spectrum of the primary component. Therefore the blue data were used to identify the spectral type of the primary, by using the prelimary identification from the earlier data and the ATLAS9 synthetic stellar template library \citep{Munari05}, rebinned as previously. We performed a least-squares fit and identified the $T_{\rm eff}=7500K$, log $g=4.0$, [Fe/H]$ = -0.5$ template as the best fit. From the residuals to this fit, we attempted to match the spectrum of the secondary component. A preliminary light curve analysis had indicated a secondary component with a temparature of $\sim$4200K; hence the least-squares fit to the residuals was restricted to the ATLAS9 models with $T_{\rm eff} = 4250K$ and [Fe/H]$ = -0.5$, since we can assume the binary system will have a common origin and thus metallicity. The best fit was achieved with the log $g=3.0$ template.

We note again that the pulsating primary component is essentially fully eclipsed by the secondary star at primary eclipse. Therefore, a high resolution spectrum during the time of primary eclipse would necessarily be a spectrum of the secondary star, and would be useful for constraining the stellar spectral type and physical parameters derived via light curve-fitting in Section \ref{sec:eb}. 

\subsection[]{Radial velocity analysis}
\label{sec:RV}

In order to extract the radial velocities of the two binary components, we used the program {\small TODCOR} \citep{Zucker94}, which performs a two-dimensional correlation between two supplied template spectra and an object spectrum of a binary system. Using the two ATLAS9 templates identified previously, radial velocities were extracted for the majority of the spectra we had obtained. The correlation was limited to the wavelength region 6200--6530~\AA~to avoid the $H\alpha$ emission noted previously. The flux ratio of the secondary to the primary template spectra was left as a free parameter for {\small TODCOR}, and was found to vary from 0.2--0.4 with phase. 

\begin{figure}
\begin{center}
\includegraphics[width=8cm]{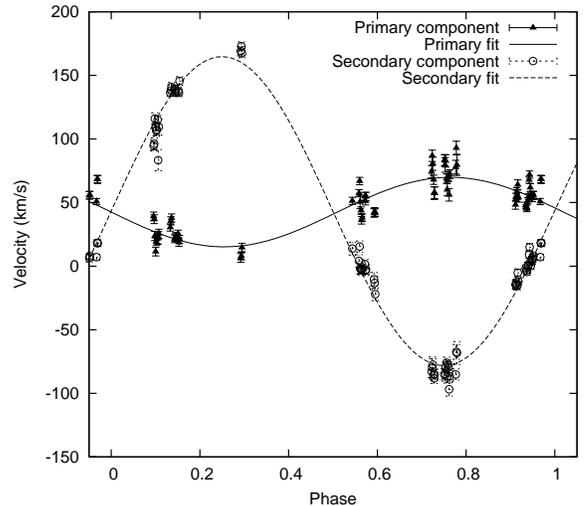}
\caption{\small The radial velocities of the primary (solid triangles) and secondary (open circles) components. The lines are the best fitting sine curves.}
\label{fig:rvs}
\end{center}
\end{figure}

The radial velocities are shown in Fig.~\ref{fig:rvs}, with the primary component shown as circles and the secondary component as squares. The lines are the best-fitting sine curves, with velocity amplitudes of $K_1 = 27.0\pm1.8$ km s$^{-1}$ and $K_2 = 121.2\pm1.4$ km s$^{-1}$, indicating a mass ratio of 0.22$\pm$0.02, and a systemic velocity of $43.6\pm0.9$ km s$^{-1}$.

The large scatter in the primary component data of $\sim$20 km s$^{-1}$ is due to the \dsc radial velocity pulsations, and is similar to the radial velocity amplitude of other HADS stars we have measured with the same instrument \citep{Derekas06}. The frequency spectrum of these data were analysed in the same manner described in Sec.~\ref{sec:puls} and two peaks corresponding to the orbital and pulsational periods (5.36 d and 0.073 d) were identified, a second confirmation that this is not a blended system.

\begin{figure*}
\begin{center}
\includegraphics[angle=-90,width=15cm]{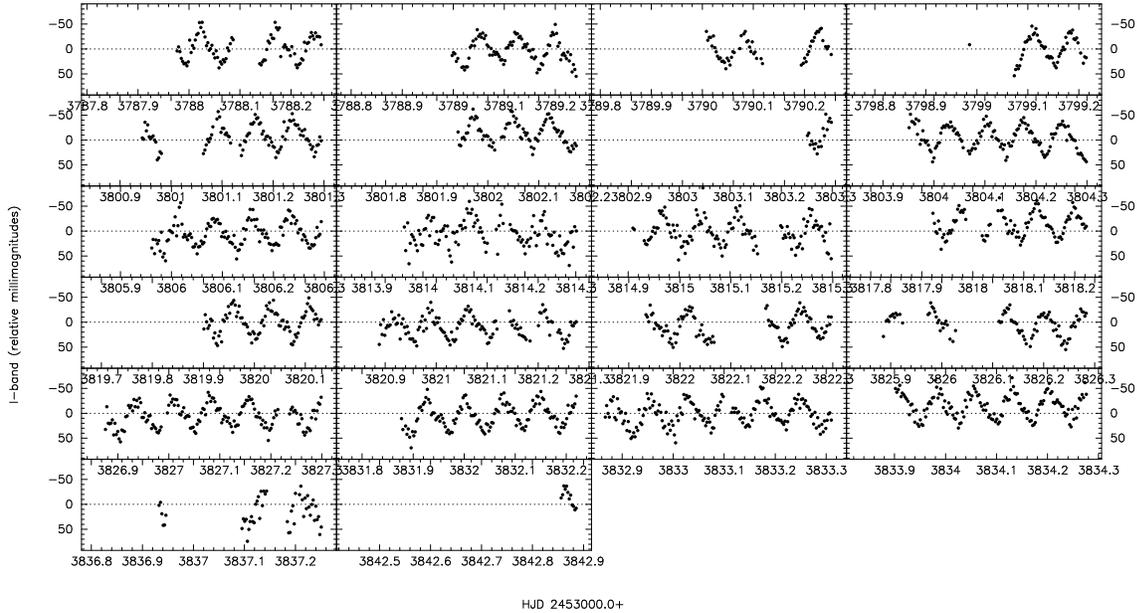}
\caption{\small The residuals in the original data after the subtraction of the best-fitting binary solution, between phases 0.1 and 0.9.}
\label{fig:res}
\end{center}
\end{figure*}

\section[]{Binary System}
\label{sec:eb}

In order to fit the orbital parameters of UNSW-V-500, we applied the Wilson-Devinney code \citep{Wilson71,Wilson79,Wilson90} to the APT light curve and the two radial velocity curves simultaneously. This system is an Algol-type semi-detached eclipsing binary system, with the secondary star filling its Roche lobe, and consequently the code was operated in mode 5. The effective temperature of the primary was fixed at $T_1 = 7500$K from the template fit. The gravity brightening coefficients were set to 1.00 for the radiative primary component and 0.32 for the convective secondary component. The albedos were set to the standard theoretical value of 1.00. The bolometric and bandpass-specific limb darkening coefficients were adopted from values for the closest models in \cite{vanHamme93}. The third light $l_3$ was assumed to be non-zero due to the crowded photometry aperture, and was allowed to vary as a free parameter. It was given an initial value of $l_3 = 0.2$ from an estimate of the maximum total contribution to the normalised flux at phase 0.25. The eccentricity was assumed to be $\sim$0 due to the secondary eclipse occurring at a phase of 0.5. To confirm this, the eccentricity was allowed to vary and did not result in any significant improvement in the fit, so was fixed at 0 for the subsequent fitting. The semi-major axis was fixed at $15.69$ \rdot~from the total mass and period of the system, and the systemic velocity was fixed at 43.6 km s$^{-1}$. The mass ratio was fixed at 0.22 from the radial velocity data. The free parameters were thus the third light $l_3$, the inclination $i$, the effective temperature of the secondary $T_2$, the potential (as defined by \cite{Kopal54}) of the primary $\Omega_1$ and the luminosity of the primary $L_1$. The values of these parameters used in the final solution are shown in Table 2, as are the derived quantities for the two components of mass, radius, log $g$ and \mbol. 

\begin{table}
 \centering
 \label{tab:wd}
\caption{The parameters for the binary system solution. Starred quantities indicate the free parameters in the Wilson-Devinney code. Quoted errors for the starred quantities and the calculated quantities ($L_2$ and $\Omega_2$) are standard deviations produced by the Wilson-Devinney code. $^a$ These are the bandpass luminosites in the $I$-band. $^b$ This is the corrected value of the third light for reference phase 0.25.}
  \begin{tabular}{@{}llllllll@{}}
  \hline
  Parameter & Value                     \\
  \hline 
  $e$                 & 0.0000          \\
  $q$*                & 0.22$\pm$0.02   \\
  $T_1$ (K)           & 7500            \\
  $T_2$* (K)          & 3850$\pm$20     \\
  $L_1$* (\ldot)$^a$             & 6.96$\pm$0.03\\
  $L_2$  (\ldot)$^a$                                                 & 3.89$\pm$0.03 \\
  $\Omega_1$*         & 6.9$\pm$0.1     \\
  $\Omega_2$          & 2.3$\pm$0.1     \\
  $i$* ($^\circ$)     & 86.5$\pm$1.0    \\
 $l_{3}$*$^b$      & 0.096$\pm$0.005 \\
  $M_1$ (\mdot)         & 1.49$\pm$0.02 \\
  $M_2$ (\mdot)         & 0.33$\pm$0.02 \\
  $R_1$ (\rdot)         & 2.35$\pm$0.02 \\
  $R_2$ (\rdot)         & 4.04$\pm$0.01 \\
  $M_{bol,1}$         & 1.80$\pm$0.02 \\
  $M_{bol,2}$         & 3.53$\pm$0.02 \\
  log $g_1$             & 3.87$\pm$0.01 \\
  log $g_2$             & 2.74$\pm$0.01 \\

  \hline
\end{tabular}
\end{table}

The solid line in Fig.~\ref{fig:lc} shows the best-fitting solution. The high inclination ($i = 86.5\pm1.0^\circ$) is indicated by the flat bottom of the primary eclipse. In fact this is the first eclipsing binary system containing a pulsating component to demonstrate a flat-bottomed eclipse, with the possible exception of the recent discovery of the pulsating component of HD 99612 \citep{Pigulski07}. The normalised third light contribution is found to be 0.096 at the reference phase of 0.25. 

We note that the log $g$ values that have been derived in the Wilson-Devinney fit ($3.87\pm0.01$ and $2.74\pm0.01$ for the primary and secondary components respectively) confirm the estimated values from the synthetic template fitting ($4.0\pm0.5$ and $3.0\pm0.5$) in Sec.~\ref{sec:spec}.

From the $T_{\rm eff}$ and derived mass we have identified the two components as a late A spectral type primary, confirming the A7V classification, and a late K spectral type secondary. Using the derived parameters, we attempted to fit the positions of the two components in the HR diagram with the Y$^2$ isochrones \citep{Yi01,Kim02b} in order to find an age estimate for the system. However, we found that the isochrones and evolutionary tracks were unable to reproduce the current positions of the components; hence we conclude that there has been significant mass transfer to the pulsating primary component from the secondary component. This component appears in a much more evolved state despite its lower mass, a well-recognized phenomenon known as the Algol paradox. We caution that standard evolution and pulsation models for HADS stars may not apply to UNSW-V-500 due to its binary evolution. However, we do note that it is well described by the ML3 mass-luminosity relation for HADS stars shown in fig. 1 of \cite{Petersen96}, which is based on models with a metal content of $Z = 0.02$. 

\section[]{Pulsation}
\label{sec:puls}

Once the binary solution has been subtracted, an analysis of the \dsc pulsation can be performed. We have used data from phase 0.1 to 0.9 for this analysis, discarding those data around the primary eclipse where the HADS star is completely eclipsed. This reduces the total number of data points by 700, or $\sim$25 per cent. The residuals from the binary subtraction between phase 0.1 to 0.9 are shown in Fig.~\ref{fig:res}.

\begin{figure}
\begin{center}
  \includegraphics[width=8cm]{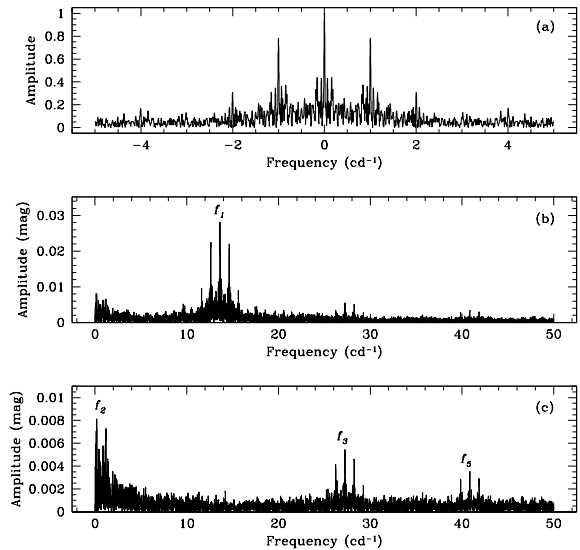}
  \caption{\small The frequency analysis of the pulsation. Panel (a) shows the spectral window. Panel (b) shows the strongest frequency $f_1$ at 13.621 cd$^{-1}$. Panel (c) shows the frequency spectrum with $f_1$ removed. On this scale, $f_4$ is coincident with $f_2$.}
  \label{fig:freq}
\end{center}
\end{figure}

Fig.~\ref{fig:freq} shows the frequency analysis, as performed with the program {\small PERIOD04} \citep{Lenz05}. The spectral window is shown in panel (a). Panel (b) shows the initial periodogram. The dominant frequency is identified as $f_1 = 13.621\pm0.015$ cd$^{-1}$, i.e. a period of $0.0734\pm0.0001$ d, which is typical for \dsc stars. The four additional frequencies identified with a S/N $>$ 4.0, as suggested by \cite{Breger93}, are shown in panel (c), after removal of $f_1$. These can be identified as low-power frequencies, probably due to artefacts of the binary subtraction ($f_2 = 0.187\pm0.033$ cd$^{-1}$ and $f_4 = 0.255\pm0.084$ cd$^{-1}$), or harmonics of $f_1$ ($f_3 = 2f_1 = 27.242\pm0.084$ cd$^{-1}$ and $f_5 = 3f_1 = 40.86\pm0.14$ cd$^{-1}$). The absence of any additional frequencies in the \dsc range supports the identification of a HADS star oscillating in a single radial mode to the limits of our detection.

As an additional check, these frequencies were removed from the original data. The resulting light curve is shown in Fig.~\ref{fig:lc}, offset below the original data. The Wilson-Devinney code was re-run using this second light curve, with no significant change in the derived parameters.

Using the relation from \cite{Breger93}, we can calculate the pulsation constant $Q_{obs}$.

\begin{equation}
{\rm log}Q_{obs} = -6.456 + {\rm log}P + 0.5{\rm log} g + 0.1M_{bol} + {\rm log}T_e
\end{equation}

We find a $Q_{obs}$ of 0.025$\pm$0.001. \cite{Petersen72} derived theoretical pulsation constants of $Q_0 = 0.0333$ for the fundamental radial mode, $Q_1 = 0.0252$ for the first overtone radial mode, and $Q_2 = 0.0201$ for the second overtone radial mode. Inspection of the HADS stars catalogued in \cite{McNamara00a} reveals that of the 26 well studied field stars, all have observed pulsation constants in the range 0.0309--0.331, indicating they are pulsating in the fundamental mode. UNSW-V-500 appears to be the first identified to be pulsating primarily in the first overtone radial mode, joining a number of SX Phe stars in globular clusters to have been identified in this mode \citep{McNamara00b}.

\section{Summary}

We have presented here the detection of the first example of a high-amplitude \dsc star in an eclipsing binary system, and the probable first detection of a field HADS star pulsating in the single first overtone radial mode. Several HADS stars have been detected in binary systems previously (including SZ Lyn \citep{Derekas03} with a period of 1190 d; and RS Gru \citep{Joner04} with a period of approximately 2 weeks), however these are much wider systems. This new fully eclipsing binary opens up many further opportunities for studies of HADS stars and pulsating stars in binary systems. Many of the poorly understood processes governing the effects of mass transfer, tidal interactions, rotation, convection and magnetism on \dsc pulsations may be explored with this system.

\section*{Acknowledgments}

The project is supported by the Australian Research Council. We gratefully acknowledge the observing time allocations from the Australian National University's Research School of Astronomy and Astrophysics. JLC is supported by an Australian Postgraduate Research Award.

\bsp

\label{lastpage}

\end{document}